\documentclass[lettersize,journal]{IEEEtran}
\usepackage{amsmath,amsfonts}
\usepackage{algorithmic}
\usepackage{algorithm}
\usepackage{array}
\usepackage{booktabs}
\usepackage[table]{xcolor}
\usepackage{diagbox} 
\usepackage{xcolor}
\usepackage{colortbl}
\usepackage[caption=false,font=normalsize,labelfont=sf,textfont=sf]{subfig}
\usepackage{textcomp}
\usepackage{stfloats}
\usepackage{url}
\usepackage{verbatim}
\usepackage{graphicx}
\usepackage{cite}
\usepackage{epstopdf}
\usepackage{pgfplots}
\usepackage{tikz}
\usepackage{hyperref}

\usepackage{times}
\usepackage{silence}
\pgfplotsset{compat=1.17}
\makeatletter

\newcommand{\Rmnum}[1]{\expandafter\@slowromancap\romannumeral #1@}
\makeatother

\begin{document}

\title{Building Low-Altitude Communication Networks: A Digital Twin-Based Optimization Framework}

\author{Boqun Huang,~\IEEEmembership{Member,~IEEE}, Yancheng Wang,~\IEEEmembership{Member,~IEEE}, Wei~Guo,~\IEEEmembership{Member,~IEEE}, Zhaojie Guo, Di Wu, Ran~Li,~\IEEEmembership{Member,~IEEE}, Dayang Liu, Wanshun Lan, Chuan~Huang,~\IEEEmembership{Member,~IEEE}, and Shuguang~Cui,~\IEEEmembership{Fellow,~IEEE} 
\thanks{B. Huang and Y. Wang are with the Shenzhen Institute for Advanced Study, University of Electronic Science and Technology of China (UESTC), the School of Science and Engineering, The Chinese University of Hong Kong, Shenzhen, China; W. Guo is with The Hong Kong University of Science and Technology, Hong Kong SAR, China; Z. Guo and D. Wu are with China Mobile Group Guangdong Co., Ltd. Shenzhen Branch; R. Li is with the Department of Information Engineering, The Chinese University of Hong Kong, Hong Kong SAR, China; D. Liu is with the China Mobile GBA Innovation Institute; W. Lan is with the China Mobile Group Guangdong Co., Ltd; C. Huang is with the Shenzhen Institute for Advanced Study, University of Electronic Science and Technology of China (UESTC), and the Shenzhen Future Network of Intelligence Institute, Shenzhen, China; S. Cui is with the School of Science and Engineering, the Shenzhen Future Network of Intelligence Institute, and the Guangdong Provincial Key Laboratory of Future Networks of Intelligence, The Chinese University of Hong Kong, Shenzhen, China.}
\thanks{Corresponding authors: Wei~Guo and Chuan~Huang (e-mail: eeweiguo@ust.hk; huangch@uestc.edu.cn).}
}

\maketitle

\begin{abstract}
Low-altitude communication networks (LACNs) serve as the critical infrastructure of the emerging low-altitude economy (LAE), supporting services such as drone delivery and infrastructure inspection. However, LACNs operate in highly dynamic three-dimensional (3D) environments characterized by high mobility and predominantly line-of-sight (LoS) propagation, creating strong coupling among key performance objectives including coverage, interference mitigation, handover management, and sensing capability. Isolated tuning of individual objectives cannot capture these cross-objective interactions, rendering conventional approaches based on experience-driven tuning and repeated field trials inefficient and costly. To address these challenges, we propose DT-MOO, a \underline{D}igital \underline{T}win-based \underline{M}ulti-\underline{O}bjective \underline{O}ptimization framework for LACNs. By constructing a high-fidelity virtual replica that integrates realistic environmental models, electromagnetic (EM) propagation, and traffic dynamics within a unified environment, DT-MOO enables joint evaluation and systematic optimization of interdependent network parameters, scoring candidate configurations by their combined effect on multiple objectives. As the foundational validation of the framework, we report real-world experiments in a 5G-enabled LACN focusing on coverage–interference co-optimization, where DT-MOO increases the high-quality coverage rate from 14.0\% to 52.9\% across all evaluated altitudes compared to an operator-provisioned, experience-based baseline, while achieving a net SINR gain under stringent criteria despite local spatial trade-offs, confirming its ability to handle coupled objectives in practical LACN deployment.

\end{abstract}


\section{Introduction}

\IEEEPARstart{T}{he} low-altitude economy (LAE) has recently attracted increasing attention worldwide and is widely expected to become an important driver of future socioeconomic growth~\cite{hosseinmotlaghLowAltitudeUnmannedAerial2016}. Uncrewed aerial vehicles (UAVs) and electric vertical takeoff and landing (eVTOL) aircraft are the key enablers of LAE, supporting applications such as logistics, infrastructure inspection, and urban air mobility. These applications require reliable communication in highly dynamic three-dimensional (3D) environments, which conventional terrestrial networks designed primarily for ground scenarios cannot adequately support~\cite{huangLowaltitudeIntelligentTransportation2024}. To bridge this gap, dedicated low-altitude communication networks (LACNs) have been proposed as the enabling communication infrastructure for low-altitude airspace~\cite{guptaSurveyImportantIssues2016,xuRecentResearchProgress2020}. However, designing LACNs that can reliably serve highly dynamic aerial users remains a significant open challenge, as the unique propagation and mobility characteristics of the low-altitude airspace give rise to a set of tightly coupled performance objectives.

Unlike terrestrial systems, LACNs operate in highly dynamic 3D environments characterized by high user mobility and predominantly line-of-sight (LoS) propagation. These characteristics create strong coupling among four key performance objectives: seamless 3D coverage, effective prediction and mitigation of bidirectional interference, reliable management of frequent handovers, and sensing capability for critical functions such as UAV detection and trajectory tracking. To illustrate this coupling, consider the example of beam width selection: wider beams reduce handovers but increase interference, while narrower beams improve interference control at the cost of coverage gaps. Optimizing any single objective in isolation is therefore insufficient and may even degrade overall system performance. Addressing these intertwined challenges demands communication infrastructures with advanced spatial resource control and integrated sensing capabilities, along with an optimization methodology capable of jointly balancing all objectives.

Among practical infrastructure candidates, satellite systems and cellular networks represent the two principal options for LACNs. Satellite communications offer wide-area coverage but suffer from large propagation delays and limited throughput, making them inadequate for LACNs. Among cellular technologies, early deployments based on fourth-generation (4G) cellular networks have demonstrated the feasibility of cellular-enabled aerial connectivity through industrial trials~\cite{QualcommTechnologiesReleases}, but 4G provides limited support for 3D beam management and sensing of high-mobility aerial targets. In contrast, fifth-generation (5G) cellular networks provide advanced beamforming and massive multiple-input multiple-output (MIMO) capabilities for finer spatial control of radio resources, while also supporting sensing functionalities essential for LACNs~\cite{xiaoSurveyMillimeterWaveBeamforming2022, songOverviewCellularISAC2025}. However, most existing studies and industrial trials~\cite{Geniusaero} still adopt isolated or sequential optimization strategies that address individual objectives such as coverage or interference alone, failing to capture the strong interdependencies among all four objectives. This raises a key question: how can these coupled objectives be evaluated and optimized in a unified and practical way?

\begin{figure*}[htbp]
\centering
\includegraphics[scale=0.63]{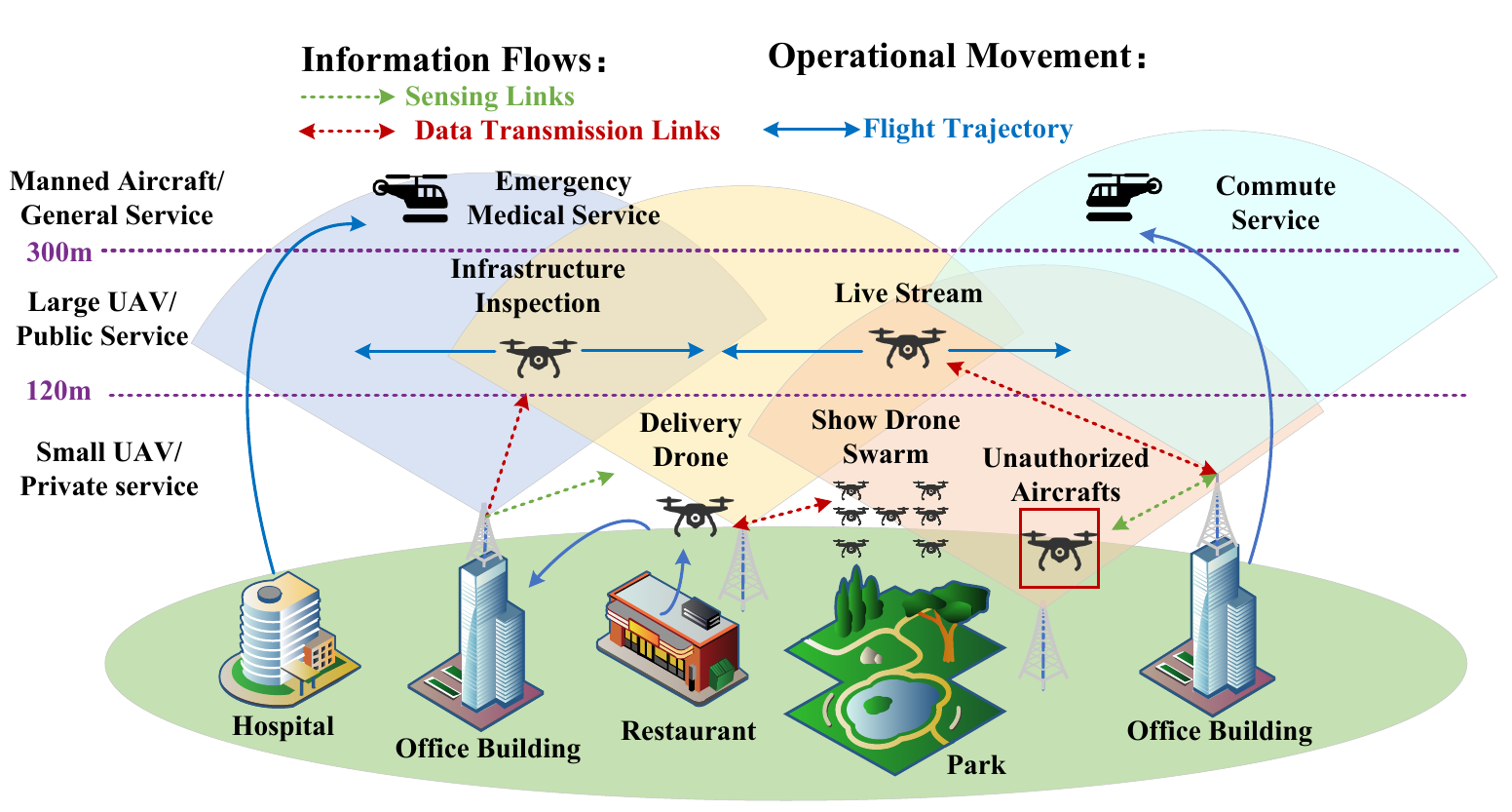}
\caption{Example LAE services categorized by altitude.}
\label{figure1}                    
\end{figure*}

To answer this question, LACNs require an evaluation platform that supports rapid, repeatable, and joint assessment of multiple coupled objectives. However, conventional simulations usually model only partial subsystems or limited metrics, and therefore cannot provide consistent system-level evaluation in a shared physical environment. Meanwhile, large-scale field experiments in low-altitude airspace are costly, difficult to repeat under controlled conditions, and impractical for iterative parameter tuning. These limitations motivate the adoption of digital twin (DT) technology. A DT establishes a high-fidelity virtual replica of the physical network by coherently integrating environmental models, electromagnetic (EM) propagation characteristics, traffic dynamics, and interference prediction, and continuously refines model fidelity through sparse field measurements. Building on this capability, we propose DT-MOO, a digital twin-based multi-objective optimization framework that enables iterative joint optimization across coverage, interference, handover, and sensing, while supporting adaptive re-optimization as network conditions change.

In this paper, we systematically analyze the coupled multi-objective challenges in 5G-enabled LACNs, elucidating the interdependencies among coverage, interference, handover, and sensing. We then develop DT-MOO, a joint optimization framework to tune key physical-layer and network-level parameters for balanced performance across multiple objectives. By leveraging the high-fidelity, field-calibrated nature of the DT, this framework substantially reduces reliance on costly and inflexible field experiments. We further present field-validated evaluations in a real LACN, including spectrum twin accuracy validation and coverage-interference co-optimization results, showing that DT-MOO improves high-quality coverage while keeping interference under control, and laying the foundation for future extensions toward broader multi-objective optimization in LACNs.

\section{Requirements and Technical Challenges for LACNs}
\IEEEpubidadjcol
This section first delineates the unique performance requirements for LACNs, spanning both communication and sensing functions, and subsequently distills four major challenges arising from these demands.

\subsection{Requirements for LACNs}
LACNs are expected to support a wide range of low-altitude services. As shown in Fig.~1, applications such as logistics, drone shows, infrastructure inspection, live streaming, and passenger transport can be grouped by operating altitude, alongside sensing tasks such as detection of unauthorized aircraft. Operating in complex 3D airspace with rapidly changing topology and dense obstacles, LACNs face much stricter requirements for communication and sensing than terrestrial networks, especially under strong air-ground interactions.

\subsubsection{Communication Requirements}
Communication requirements in LACNs typically evolve from basic network-level coverage to more stringent route-level service guarantees. At the most fundamental level, the network must provide continuous connectivity throughout irregular 3D airspace. The propagation environment varies significantly with altitude: LoS propagation dominates at higher altitudes, while operations near urban structures still experience multipath effects. The network must therefore cope with altitude-dependent propagation conditions as well as Doppler shifts induced by aerial mobility. On top of this basic capability, mission-critical services along specific flight routes, such as infrastructure inspection or emergency response, require higher communication quality. These services often demand high uplink throughput for real-time video transmission, low latency on the order of tens of milliseconds for stable control, and very high reliability to accommodate rapid mobility and channel fluctuations.

\subsubsection{Sensing Requirements}
Sensing requirements also evolve from device-level perception capability to system-level communication-sensing integration. At the device level, safe operation in cluttered urban environments requires high-precision and real-time 3D perception to track heterogeneous aerial traffic and avoid collisions even under non-line-of-sight (NLoS) conditions. At the network level, sensing is integrated with communication within the integrated sensing and communication (ISAC) framework, where sensing and communication share radio resources. This integration requires careful coordination to control co-channel interference and a robust system architecture capable of maintaining stable operation under dense traffic conditions and potential cyber threats \cite{wuComprehensiveOverview5GandBeyond2021}.

\begin{figure*}[htbp]
\centering
\includegraphics[scale=0.76]{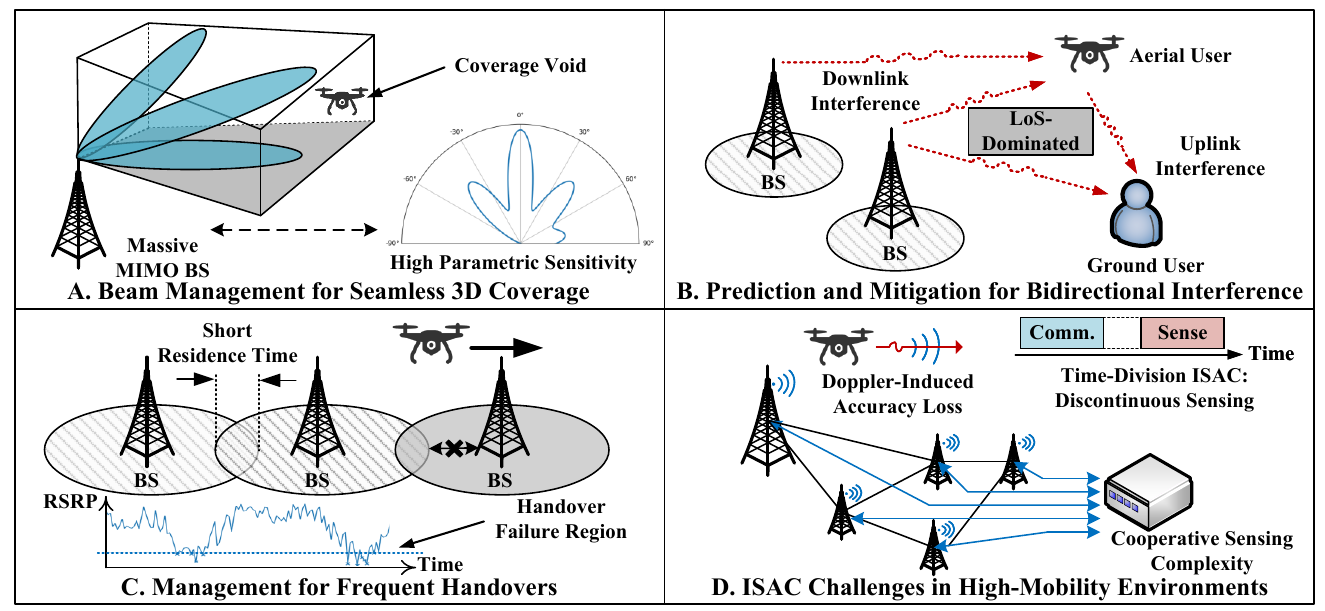}
\caption{Technical challenges of 5G for LACNs.}
\label{figure2}                    
\end{figure*}

\subsection{Technical Challenges of 5G for LACNs}
As discussed in Section I, 5G is widely regarded as the most promising candidate for LACNs. However, applying it to low-altitude scenarios is not straightforward. This section summarizes several key challenges, as shown in Fig.~2.

\subsubsection{Beam Management for Seamless 3D Coverage}
To illuminate the vertically extended airspace, BSs naturally employ upward-tilted beams, which introduces two effects absent in 2D terrestrial deployments. First, signals propagate over much longer distances under near-LoS conditions, reshaping the spatial footprint of each cell and easily leaking energy into neighboring areas. Second, each BS leaves a pronounced coverage hole directly above itself, which must be filled by upward-tilted beams from neighboring BSs through careful inter-cell coordination. Although 5G massive MIMO provides sufficient spatial degrees of freedom, exploiting them is non-trivial, as a small adjustment in beam tilt or width may unintentionally create coverage gaps elsewhere or disrupt the inter-cell coordination required~\cite{wang2025precisechannelknowledgemap}.

\subsubsection{Prediction and Mitigation for Bidirectional Interference}
The elevated and near-LoS propagation paths in LACNs also reshape the interference landscape. Aerial users suffer strong multi-cell downlink interference from beam leakage of distant BSs, while their uplink transmissions propagate freely to interfere with terrestrial users served by surrounding BSs. This LoS-induced coupling tightly binds interference to beam configuration, so interference mitigation must be considered jointly with coverage optimization rather than as an isolated objective.

\subsubsection{Management for Frequent Handovers}
The high speed of flying platforms such as UAVs shortens cell residence time, leading to frequent handover attempts and increased failure risk. The aerial channel also varies rapidly with altitude, orientation, and trajectory, causing large fluctuations in reference signal received power (RSRP) that make conventional signal-strength-based handover criteria unreliable. Designing reliable handover prediction and decision mechanisms along mission routes therefore remains a key challenge.

\subsubsection{ISAC Challenges in High-Mobility Environments}
While 5G-Advanced enables BSs to perform ISAC, applying it to LACNs is non-trivial~\cite{mengUAVEnabledIntegratedSensing2024}: high mobility aggravates the limitations of single-node and time-division schemes, leading to interrupted sensing continuity, reduced spectrum efficiency, and Doppler-related accuracy degradation. Addressing these issues requires cooperative sensing across multiple nodes and jointly designed ISAC waveforms, whose joint evaluation and optimization under strict LACN requirements remain a key challenge.

Taken together, these tightly coupled challenges form a complex optimization problem that cannot be effectively addressed by experience-based tuning or isolated analytical models~\cite{mozaffari6GConnectedSky2021}, motivating the DT-MOO framework proposed in the next section.

\section{DT-MOO Framework}

As discussed in the previous section, LACNs require a unified framework for network modeling and optimization. DT technology offers a promising solution: it provides a continuously updated virtual representation of the physical network by integrating data, models, system interfaces, and real measurements~\cite{taoDigitalTwinIndustry2019}. For LACNs, DT supports joint evaluation of coupled objectives such as 3D coverage, interference, handover, and sensing, remains synchronized with dynamic network conditions, and enables pre-deployment testing with reduced aerial measurement cost. Motivated by these advantages, we develop the DT-MOO framework for multi-objective optimization in LACNs.

\begin{figure*}[htbp]
\centering
\includegraphics[scale=1]{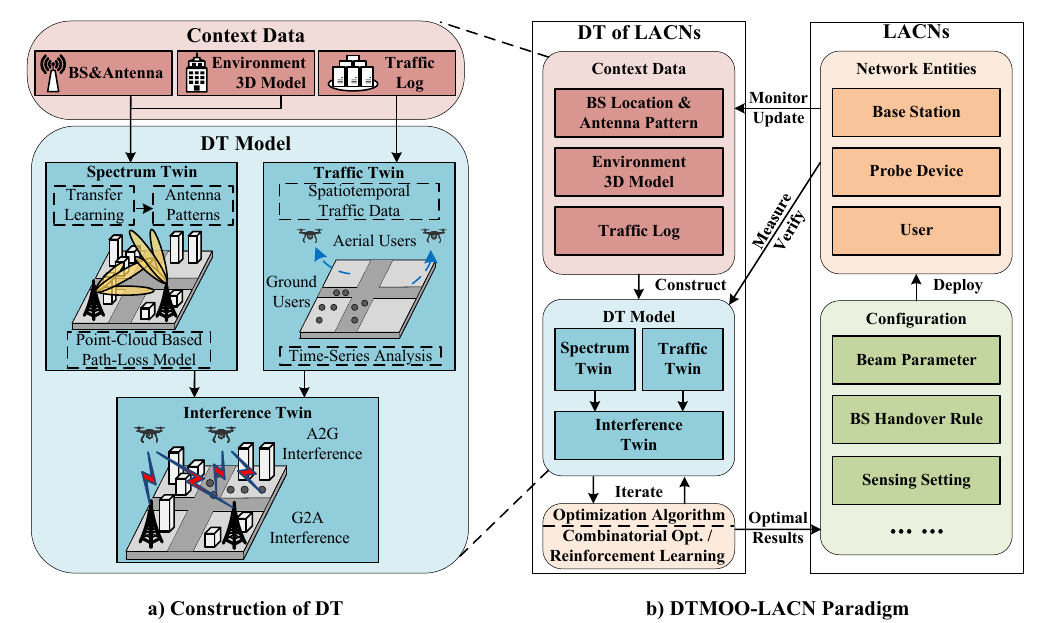}
\caption{DT of LACNs and DT-MOO workflow.}
\label{figure3}                    
\end{figure*}

\subsection{DT Construction for LACNs}
The DT used in this work replicates the key elements of a physical LACN. It consists of three tightly connected modules: the spectrum twin, the traffic twin, and the interference twin. As illustrated in Fig.~3(a), the spectrum twin models the radio propagation environment and provides the basis for coverage analysis and beam management. The traffic twin represents the spatial and temporal distribution of network demand, which reflects user mobility and service requirements. The interference twin combines information from the spectrum and traffic twins to estimate interference levels across the network. Together, these modules allow the DT to support system-level analysis of functions such as handover management and ISAC.

\subsubsection{Spectrum Twin}
The spectrum twin models the radio field by combining EM propagation models, realistic antenna radiation patterns, and detailed environmental information. Among these elements, accurate antenna pattern modeling is the common cornerstone for spectrum twin construction in LACNs. Because LACNs cover a large vertical space and rely on upward-tilted beams to illuminate the airspace, even small errors in the antenna pattern can lead to noticeable prediction errors, and these errors tend to grow with altitude. Moreover, antenna patterns at different pointing angles cannot be obtained simply by rotating a single template; in practice, accurate patterns are measured in anechoic chambers, where each configuration must be tested individually, making exhaustive measurement prohibitively expensive. To address this issue, we develop a transfer-learning-based method to reconstruct antenna radiation patterns at arbitrary pointing angles. The antenna model is first trained on EM simulation data and then refined using a small set of measured patterns obtained from anechoic-chamber experiments. The resulting antenna model serves as a shared building block for the propagation modeling described below.

Built upon this antenna foundation, the spectrum twin adopts two propagation modeling schemes tailored to different low-altitude environments. In open low-altitude environments dominated by LoS propagation, the radio field is reconstructed by combining a free-space path loss (FSPL) model with the learned antenna pattern, which is sufficient to capture the dominant propagation behavior in sparsely scattering scenarios. In dense urban airspace, where multipath propagation and blockage become more significant, we further adopt a point-cloud-based method together with the same antenna model~\cite{wangChannelKnowledgeMaps2024}. Arrival-time information is first used to construct transmitter-receiver co-focal ellipsoids that indicate possible reflection and scattering regions. An ellipsoid-based filtering method is then applied to point-cloud data to identify potential reflection, scattering, and attenuation points. A joint data- and model-driven learning approach is used to map these filtered points to sparse spectrum samples~\cite{wang2025pointcloudenvironmentbasedchannel}. By incorporating propagation principles and antenna directionality, the model estimates path loss with explicit consideration of the angle of departure. Multipath components are then grouped according to spatial or angular similarity, and the power of each component is calculated from the estimated path loss and antenna gain. These components are aggregated to estimate the received power at arbitrary 3D locations. The resulting radio field forms the spectrum twin and provides the EM foundation for further analysis.

\subsubsection{Traffic Twin}
The traffic twin describes the distribution of user demand in the low-altitude network. It is constructed using historical data and continuously updated with real-time information. To achieve this, we integrate multiple data sources, including service records from aerial and ground users, real-time traffic measurements, and operational airspace information such as geofences, restricted zones, and planned flight routes.

A time-series analysis model is used to capture temporal trends and spatial correlations in network usage. Based on this model, the spatiotemporal distribution of user demand can be estimated. This information is embedded into the spatial radio environment described by the spectrum twin, allowing the DT to represent the dynamic distribution of traffic load across the network.

\subsubsection{Interference Twin}
Based on the propagation environment and traffic demand, an interference twin is constructed to analyze bidirectional interference between aerial and ground users. The interference model considers both ground-to-air and air-to-ground interference.

Ground-to-air interference is estimated using the spectrum twin, which predicts both direct and, in dense settings, reflected signal components from BSs. These predictions allow the distribution of interference affecting aerial links to be evaluated. Air-to-ground interference is modeled by dividing the ground network into spatial grids and estimating the interference generated by UAV transmissions. The grid resolution can be adjusted according to service requirements, and higher resolution is used in areas with dense population or near important flight routes and takeoff or landing locations. This approach provides a detailed view of interference conditions across the network.

\subsection{Workflow of DT-MOO}
Based on the constructed DT of LACNs, we develop a DT-MOO framework to jointly optimize multiple performance dimensions of the network. As illustrated in Fig.~3(b), the overall workflow is organized into three phases: DT Initialization, DT Optimization, and DT Monitoring and Adaptation, forming a closed loop that connects data acquisition, virtual modeling, algorithmic optimization, and physical deployment.

\textbf{1. DT Initialization.}
The process begins with the initialization of the DT, which serves as the foundational virtual representation of the physical LACN. This phase involves aggregating a diverse set of real-world inputs, including historical traffic data, BS configurations, and antenna patterns. These static inputs are further augmented by 3D environmental models, point-cloud data of the terrain, and EM principles to accurately model the propagation environment. These heterogeneous data streams are then synthesized to construct a high-fidelity DT composed of three tightly coupled modules: the spectrum twin, the traffic twin, and the interference twin.

\textbf{2. DT-Based Optimization.}
Following initialization, the DT enters an optimization phase, operating as a virtual sandbox for enhancing the network. DT-MOO formulates LACN optimization as a multi-objective problem in which multiple key performance indicators (KPIs), such as RSRP, signal-to-interference-plus-noise ratio (SINR), throughput, and latency, are jointly evaluated within the same DT environment, so that trade-offs among them are captured within a single optimization loop. Based on this unified evaluation, an optimization module iteratively tunes network parameters to find superior configurations. Depending on the network scale and objective complexity, two categories of algorithms are employed: combinatorial optimization for fast, near-optimal solutions, and reinforcement learning for global policy exploration. The tunable parameters span two levels: at the BS level, per-beam configuration parameters such as direction (azimuth and elevation) and pattern shape; at the system level, power allocation across beams, handover rules, and sensing parameters like waveform selection and the constant false alarm rate. Crucially, this iterative evaluation-optimization loop occurs entirely within the virtual domain, enabling the risk-free exploration of a vast solution space, which would be impossible in physical field trials due to safety and cost constraints.

\textbf{3. DT Monitoring and Adaptation.}
Once the optimized configuration is deployed to the physical LACN, the workflow transitions to the last phase, focused on maintaining synchronization and enabling continuous adaptation. This phase begins with cross-validation, where initial field measurements are collected and compared against DT predictions to ensure model fidelity. Subsequently, a continuous monitoring process is activated. BSs, user terminals, and dedicated probe devices collect dynamic data, including real-time traffic loads, spectrum measurements, and low-altitude flight trajectories.

This constant influx of real-time data is fed back to the DT to maintain virtual-physical synchronization, ensuring the twin accurately reflects the current state of the physical LACN. Once the user requirements change, for instance, if the user's flight route changes, the DT-MOO workflow can promptly provide corresponding LACN parameter optimization suggestions.

\section{Experiments}
As the foundational layer of DT-MOO, this paper validates the spectrum twin fidelity and the coverage-interference co-optimization it directly enables, leaving handover and ISAC optimization to future work. For this purpose, we established a real-world LACN testbed. Unlike terrestrial deployments, this site utilizes six 5G BS cells with upward-tilted antennas to illuminate a cylindrical airspace (2 km radius, 0-500 m altitude). We employed a DJI Matrice 300 RTK UAV equipped with an R\&S TSMA6 scanner to capture high-precision measurements, discretizing the airspace into $10 \times 10 \times 10$ m³ voxels.

To evaluate performance, we define effective service coverage based on a joint analysis of RSRP and SINR. This coupled criterion is critical in dense networks, where increasing transmit power to improve RSRP often degrades SINR due to co-channel interference. Accordingly, a voxel is considered covered only if it simultaneously satisfies the thresholds for signal strength (e.g., $-95$ dBm) and signal quality (e.g., $-3$ dB), ensuring a robust communication link. The baseline BS configuration was provided by the network operator and tailored to low-altitude communication requirements based on practical deployment experience and feedback from on-site measurements. All comparisons between this baseline and the optimized configuration were conducted using identical UAV flight routes to ensure fairness. To foster reproducibility and facilitate further research in the LACN community, we have open-sourced the measurement dataset on GitHub\footnote{The dataset is available at: \url{https://github.com/ycw671/LACN-Dataset}}.

\subsection{DT-MOO Implementation}
The application of the DT-MOO framework in the experimental field proceeds in three stages: construction, optimization, and maintenance.

\begin{enumerate}
\item \textbf{Spectrum Twin Construction:} Since the test area is located in mountainous terrain with sparse building blockage, the channel is dominated by LoS propagation. Consequently, we constructed the spectrum twin by combining a FSPL model with the learned antenna pattern described in Section III. This approach provides a high-fidelity, voxel-level simulation environment capable of accurately predicting signal propagation and interference patterns.
\item \textbf{Joint Optimization of Coverage and Interference:} We formulated the optimization of the 42 configurable sub-beams (seven per cell across six cells) as a multi-objective combinatorial problem that jointly accounts for coverage and inter-cell interference. To achieve efficient solutions, we employed a DT-assisted greedy sequential selection strategy. For each sub-beam, the algorithm evaluates candidate steering angles via the spectrum twin, scoring each candidate by its joint contribution to both the RSRP improvement over the target airspace and the SINR margin in neighboring voxels. To further reduce inter-cell interference, an early-termination rule is incorporated: when the marginal gain of a new steering angle becomes small, the algorithm reuses an angle already selected for another sub-beam in the same cell, naturally aligning sub-beams within each cell.
\item \textbf{Lifecycle Management:} To maintain long-term utility, the framework integrates a continuous monitoring loop in which empirical data from BSs and UAV scans are compared against DT predictions, with significant discrepancies triggering an adaptation cycle to fine-tune underlying models such as antenna patterns.
\end{enumerate}

\subsection{Spectrum Twin Validation}
Before deploying the optimization strategy, we first verify the fidelity of the spectrum twin, as accurate radio propagation reproduction is fundamental to our DT-based framework.

\begin{figure*}[htbp]
    \centering
    \captionsetup[subfigure]{font=scriptsize,labelfont=rm,textfont=rm} 
    \subfloat[Absolute prediction errors along the UAV trajectory]{
        \includegraphics[width=0.5\linewidth]{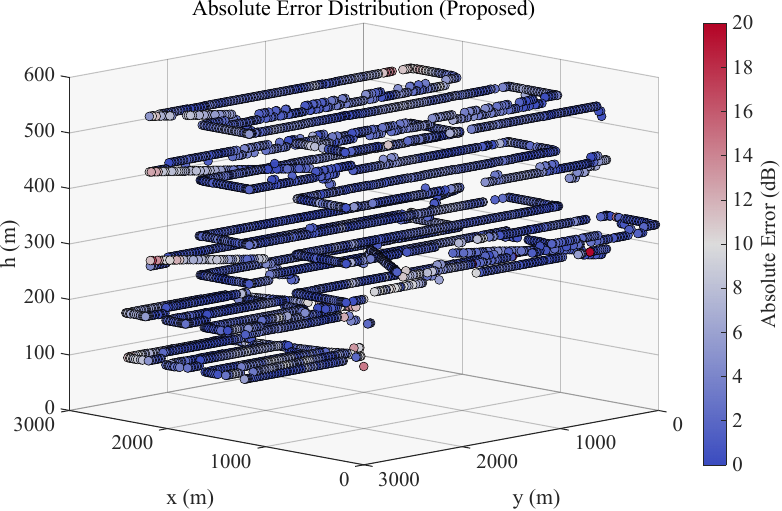}
        \label{fig:4-1}
    }
    \hfill
    \subfloat[CDF of prediction error for RT, Kriging, and the proposed method]{
        \includegraphics[width=0.45\linewidth]{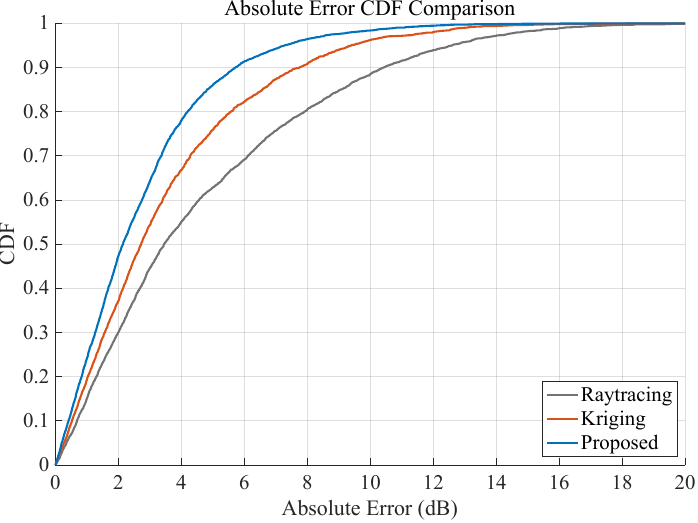}
        \label{fig:4-2}
    }
    \caption{Validation of the proposed spectrum twin.}
    \label{fig:comparison}
\end{figure*}

To evaluate the prediction accuracy, the proposed spectrum twin is compared with two representative baselines: a conventional RT propagation model and a data-driven Kriging interpolation method. The RT baseline reconstructs multipath components from 3D environmental geometry and material parameters, while the Kriging baseline interpolates the measured data in a height-wise manner, treating each altitude layer as an independent 2D interpolation problem. All three methods are evaluated under the same 70-30 split protocol, in which 70\% of the samples along the UAV trajectory are used for training and the remaining 30\% form a contiguous test segment, shifted along the trajectory across three folds under a block hold-out scheme. What differs is how each method uses the 70\%: the RT and proposed models use it only to calibrate a single global offset absorbing deployment-dependent factors such as BS transmit power and receiver antenna gain, while Kriging uses the full 70\% as interpolation nodes. All methods are then evaluated on the same test locations to ensure a fair comparison.

Fig.~4(a) visualizes the absolute prediction errors of the proposed spectrum twin along the UAV trajectory, where blue indicates small errors and red indicates larger deviations. Most trajectory points remain within the low-error range across different positions and altitudes, demonstrating that the spectrum twin captures the variation of radio signals in the LACN.

\begin{figure*}[htbp]
\centering
\includegraphics[width=\textwidth]{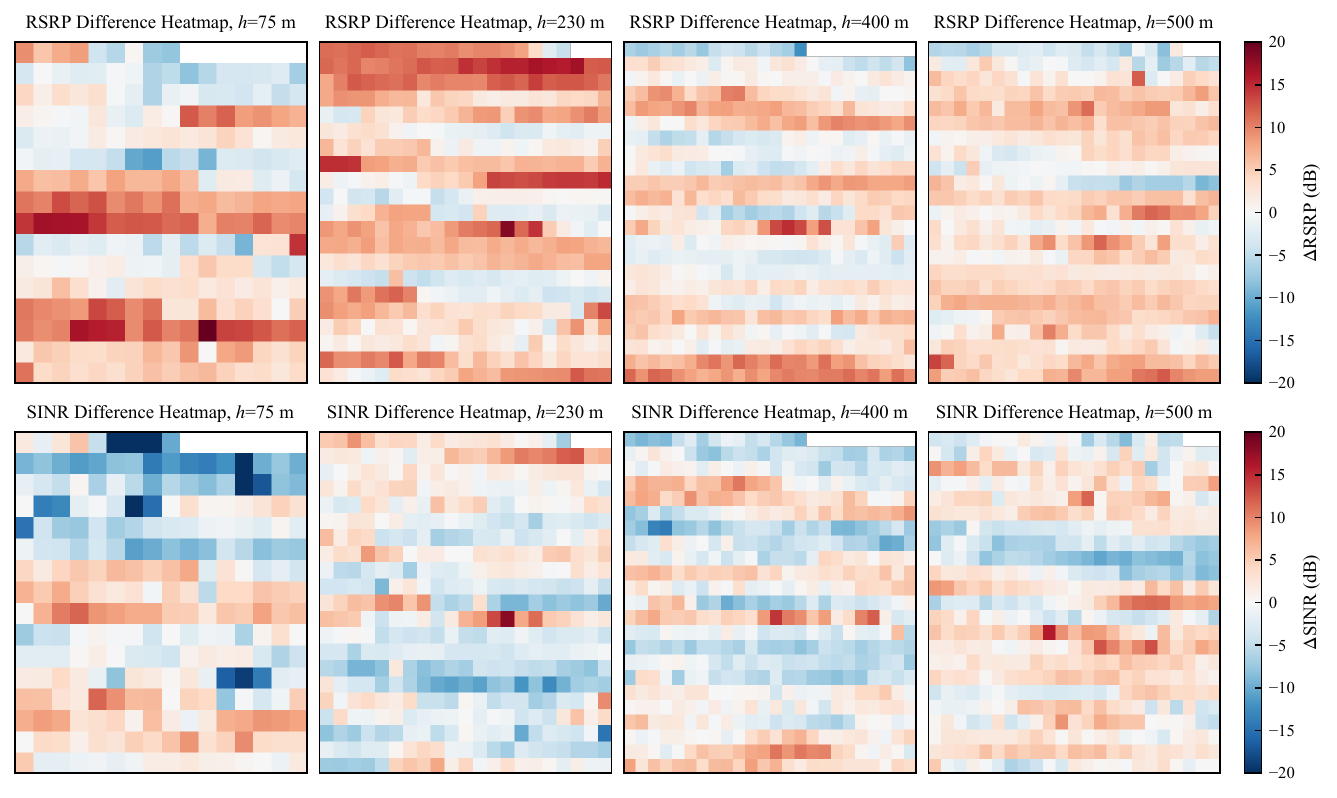}
\caption{Spatial heatmap comparison of RSRP and SINR before and after DT-MOO optimization}
\label{figure5}                    
\end{figure*}

Fig.~4(b) presents the cumulative distribution function (CDF) of the absolute prediction errors for the three methods. The proposed spectrum twin consistently achieves lower prediction errors than both baselines. Quantitatively, the RT model yields an RMSE of 6.17~dB and the Kriging interpolation achieves 4.67~dB, while the proposed spectrum twin achieves a significantly lower RMSE of 3.66~dB, corresponding to a 40.7\% reduction relative to RT and a 21.6\% reduction relative to Kriging.
The proposed spectrum twin outperforms the RT baseline likely because the considered low-altitude scenario is LoS-dominated, where the received power is mainly determined by the direct path. Since both methods use the same antenna radiation pattern, the performance gap mainly comes from the propagation model. In this case, the additional multipath components modeled by RT contribute little to prediction accuracy, while errors in the input 3D geometry, assumed material properties, and coherent multipath combining may degrade its accuracy. As a result, an FSPL-based model combined with a high-fidelity antenna pattern can outperform a more sophisticated RT-based model in the considered scenario. However, it is worth noting that RT remains a powerful choice in scenarios where reflection, diffraction, and other multipath effects are non-negligible.

\subsection{Optimization Results}
Having established the fidelity of the spectrum twin, we then evaluated the optimization performance. The core challenge in LACN joint optimization of coverage and interference is the strongly coupled nature of the objectives: maximizing coverage often inadvertently increases inter-cell interference, thereby degrading signal quality. Our experimental results demonstrate how the DT-MOO framework effectively navigates this trade-off.

\begin{figure}[htbp]
\centering
\includegraphics[width=\columnwidth]{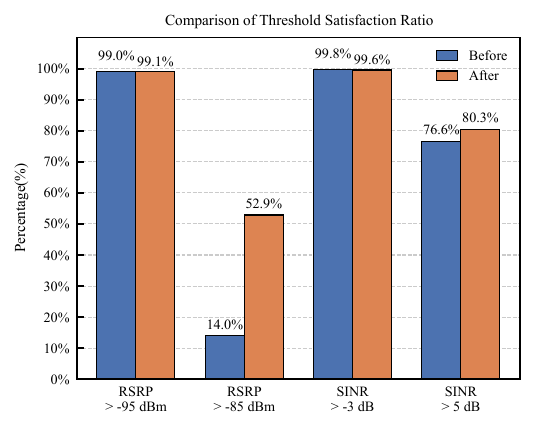}
\caption{Threshold satisfaction ratios of RSRP and SINR before and after DT-MOO optimization}
\label{figure6}
\end{figure}

Fig.~5 provides a spatial view of the optimization outcomes through difference heatmaps before and after DT-MOO at representative altitudes. The RSRP difference maps show a broad and consistent enhancement of signal strength across the airspace, indicating that many previously weak regions are upgraded to stronger coverage conditions. The SINR difference maps exhibit a more nuanced pattern: at higher altitudes SINR is broadly improved, while at lower altitudes the substantial coverage enhancement in certain regions propagates as inter-cell interference into their neighbors, causing localized SINR reductions. This spatial trade-off reflects the strongly coupled nature of coverage and interference, and shows that DT-MOO does not blindly maximize RSRP but redistributes radio resources across 3D space in a coordinated manner. Whether the net effect on SINR is favorable is further quantified in Fig.~6.

To quantify whether these spatial trade-offs lead to a net improvement, Fig.~6 reports the threshold satisfaction ratios of RSRP and SINR aggregated across all height levels. For RSRP, DT-MOO improves the ratio from 14.0\% to 52.9\% under the stricter threshold ($>-85$ dBm), while maintaining near-complete coverage under the basic threshold ($>-95$ dBm). For SINR, the ratio under the stricter threshold ($>5$ dB) also improves from 76.6\% to 80.3\%, while the basic threshold ($>-3$ dB) remains high before and after optimization. 

These aggregated results confirm that the local SINR variations observed in Fig.~5 are outweighed by gains elsewhere in the airspace, yielding simultaneous net improvements in RSRP and SINR under stringent criteria. By internalizing this coupled relationship within the DT, DT-MOO enables practical, data-driven parameter tuning with substantially reduced dependence on repeated field trials.

\section{Conclusion and Future Work}
LACNs present a unique set of intertwined challenges, including beam management for seamless 3D coverage, the prediction and mitigation of bidirectional interference, and the complexities of handover management and ISAC in high-mobility environments. The strong interdependencies among these factors render traditional optimization methods inadequate, as they rely on costly physical measurements and are often limited to single-objective optimization.

To address these limitations, this paper proposes DT-MOO, a novel framework for multi-objective optimization in LACNs. At its core, this framework leverages a high-fidelity DT, constructed from a spectrum twin, a traffic twin, and an interference twin, to migrate the network optimization process from the physical environment to a virtual DT environment.

Real-world experiments validated DT-MOO on the representative case of coverage-interference co-optimization, showing that coordinated radio resource redistribution across 3D space substantially improved coverage quality while achieving a net SINR gain under stringent criteria.

\begin{itemize}
    \item \textbf{Extended ISAC and multi-modal environmental perception:}  
    While this framework focuses on radio-based ISAC, future research can extend DT-MOO to support richer environmental perception by incorporating multi-modal sensing information, such as vision, radar, and infrared. A DT provides a natural platform to fuse heterogeneous sensing modalities and jointly optimize communication, sensing, and perception tasks under unified performance objectives. This extension is particularly relevant for addressing sensing blind spots, interference-limited scenarios, and safety-critical low-altitude applications.

    \item \textbf{Airspace-network co-optimization and route-aware planning:}  
    Beyond coverage optimization, DT-MOO can be extended to jointly consider low-altitude airspace utilization and communication network performance. By integrating service-specific flight patterns, airspace constraints, and traffic dynamics into the DT, future work can enable route-aware optimization that accounts for communication quality, sensing reliability, latency, and interference along candidate flight corridors. 

    \item \textbf{Mobility-aware and knowledge-driven optimization:}  
    High mobility in low-altitude environments introduces additional challenges, such as Doppler effects, rapidly varying channels, and frequent topology changes. Future work can enrich DT-MOO with mobility-aware models and knowledge-driven representations, such as channel knowledge maps (CKMs), to better capture spatiotemporal channel evolution. By continuously updating these models through real-time measurements, the DT can support predictive and proactive optimization strategies that improve robustness under fast-varying aerial dynamics.
\end{itemize}

\bibliographystyle{IEEEtran} 
\bibliography{IEEEabrv,ref}

\section*{Biography}

\noindent\textbf{Boqun Huang} is a Ph.D. student in the School of Science and Engineering at The Chinese University of Hong Kong, Shenzhen, China.

\vspace{2mm}
\noindent\textbf{Yancheng Wang} is a Ph.D. candidate in the School of Science and Engineering at The Chinese University of Hong Kong, Shenzhen, China.

\vspace{2mm}
\noindent \textbf{Wei Guo} received the Ph.D. degree from The Chinese University of Hong Kong, Shenzhen. He is currently a Postdoc researcher at The Hong Kong University of Science and Technology.

\vspace{2mm}
\noindent \textbf{Zhaojie Guo} received the Master's degree from Sun Yat-sen University, Guangzhou, China, in 2007. He is currently a Communications Engineer and a Provincial Expert with China Mobile Group Guangdong Co., Ltd. Shenzhen Company.

\vspace{2mm}
\noindent \textbf{Di Wu} received the Bachelor's degree from the University of Electronic Science and Technology of China, Chengdu, China, in 2007. He is currently a 5G Technical Supervisor in the Network Department and a Provincial Expert with China Mobile Group Guangdong Co., Ltd. Shenzhen Branch.

\vspace{2mm}
\noindent\textbf{Ran Li} received the Ph.D. degree in computer and information engineering from the Chinese University of Hong Kong, Shenzhen, China, in 2023. He is currently a Research Associate with the Chinese University of Hong Kong, Hong Kong, China.

\vspace{2mm}
\noindent\textbf{Dayang Liu} received his B.S. degree in Communication Engineering from the University of Electronic Science and Technology of China in 2008, and his M.S. degree in Electronic and Communication Engineering from Sun Yat-sen University in 2015.
He is a Deputy General Manager of CMCC Guangdong Bay Area Innovation Research Institute Co., Ltd. He is also a doctoral supervisor at the School of Electronic and Information Engineering (School of Microelectronics), Sun Yat-sen University, and a Standing Director of the Guangdong Institute of Communications. 

\vspace{2mm}
\noindent\textbf{Wanshun Lan} received his B.S. degree in electronic engineering from Beijing University of Posts and Telecommunications, China, in 2000. He is currently a Distinguished Expert of the Academic Committee of the Key Laboratory of Short-Range Radio Device Detection and Assessment, Ministry of Industry and Information Technology, a member of the Guangdong Provincial Senior Professional Title Evaluation Committee for Information and Communication Engineering Technical Personnel, and a member of the Guangdong Provincial Technical Committee on Artificial Intelligence Standardization.

\vspace{2mm}
\noindent\textbf{Chuan Huang} received the Ph.D. degree from Texas A\&M University, College Station, TX, USA. He is currently a Professor with the Shenzhen Institute for Advanced Study, University of Electronic Science and Technology of China, and the Shenzhen Future Network of Intelligence Institute.

\vspace{2mm}
\noindent \textbf{Shuguang Cui} received the Ph.D. degree from Stanford University, Stanford, CA, USA. He is currently the X.Q. Deng Presidential Chair Professor in the School of Science and Engineering at The Chinese University of Hong Kong, Shenzhen, China.

\vfill

\end{document}